# Nanomechanics of single keratin fibres: A Raman study of the α helix → β sheet transition and water effect

Raphaël Paquin [a], Philippe Colomban [a] *

Laboratoire Dynamique, Intéractions et Réactivités, UMR 7075 CNRS – Université Pierre et Marie Curie-Paris 6, 2 rue Henry-Dunant, 94320 Thiais, France

*fax 33 1 4978 1118
philippe.colomban@glvt-cnrs.fr

## ABSTRACT

The use of micro-Raman spectroscopy, through chemical bond nano-scale probes, allows the changes in conformations (α helix → β sheet), chain orientation, disconnection of disulfide bonds (-20%) and the increase of intra and inter-chain distances during the strain to be distinguished. The combination of micro-Raman spectroscopy and a allows a quantitative measure of the extension of chemical bonds in the peptidic chain during loading. The nano-structural transformations of keratin during the strain of human hair in a dry environment (40-60 % relative humidity) and saturated with water have been studied. The water permits the sliding of the chains and decreases the bond energy hair. Spectral analyses and 2D correlation are two coherent and independent methods to follow change the Raman probes which are sensitive to structural . The between nano-mechanical (Raman) and micro-mechanical (strain/stress) analyses confirms the validity of the experimental results, tools and principles used, as well as the agreement with the structural model of keratin fibres described by Chapman & Hearle.

The micro-Raman spectroscopy using chemical bond nano-scale probes allowsthe changes in conformations (α helix → β sheet), chain orientation, breakage of disulfide bonds (-20%), interaction with water and the increase of intra and inter-chain distances during the strain to be distinguished. Results are in agreement with the structural model of keratin fibres described by Chapman & Hearle.

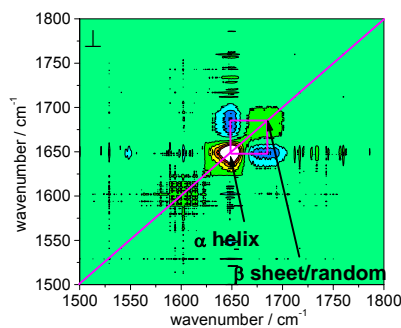

**Nanomechanics of single keratin fibres: a Raman study of the α helix → β sheet transition and water effect**

*Raphaël Paquin and Philippe Colomban*

# INTRODUCTION

The structural mechanics of keratin fibres have been recently reviewed [1,2]. Like wool, hair fibres have a hierarchical structure, sketched in Figure 1. They consist of two basic components: the cuticle and the cortex. The cuticle consists of a scale structure over the surface of the cortex. Hence the central part: cortex ($\varnothing = 80$ µm), is fully formed by elongated cells of an irregular cross section of few micrometers (Feughelman, 2002). Under an electron microscopy, a composite structure is revealed, macrofibrils ($\varnothing = 0.4$ µm) are seen to compose the sub-structure of cortical cells and to pack together intermediate filaments (IFs, $\varnothing = 8$ nm) [3]. IFs are assumed to be embedded in a rich cystine amorphous matrix. More recently, X-Ray Diffraction studies have given an insight into the architecture of the Ifs at a shorter scale, constituting of α-keratin helix dimers (coiled coils, $\varnothing = 1$ nm) assembled most likely in two tetrameric oligomers [4]. Keratotic proteins consist of heterogeneous amino-acids chain [5], and to assembly keratin macromolecules together, two types of interaction must exist: strong disulphides (–S-S-) and weak hydrogen bonds (–N-H···O).
Location of the –S-S- bonds (IFs or matrix/IFs) [1,6] are however at the origin of the divergence of the structural mechanics models. So far, in the literature, no mechanical contribution of the disulphides bonds has been reported using vibrational spectroscopy. X-Ray diffraction studies have revealed that the effects of water must involve a combination of a sliding and molecular stretching process [7]. Indeed, at low humidity, it is well-know that the α helix → β sheet transition occurs as a consequence of the macroscopic stress [8-10].

Analogy with less complex organized materials, such as polymer fibres, shall be considered in this study. As demonstrated for different synthetic polymer fibres (PET, PA-66, i-PP) examined by Raman spectroscopy [11-14], interactions, chemical bonds strength, and macromolecules rearrangements determine the macromechanical properties (Fig. 2). A correlation is obvious between the fibre's density (number of chemical bond per volume unit) and the ultimate stress. Thanks to Raman studies on the cosmetic effect [15-18], ageing of hair [19], or more generally Raman analysis on keratotic materials (hair, wool, skin, nail) [20-22], Raman modes are available to examine interchain arrangements and macromolecular conformation.The analyse of the deformation of hair fibres in dry air and in water using Raman spectroscopy has been carried out in this study in order to investigate the role of water in the mechanics of keratin fibres. the behaviour of the –S-S bonds will be paid with the aim of clarifying ambiguous claims in the literature.

# EXPERIMENTAL

### Materials
Virgin white hair, free of any cosmetic action, to be used as keratin fibres was taken from a Caucasian subject (average fibre diameter: 80 µm), so as to prevent fluorescence or sample destruction due to laser exposure: white hair do not contain melanin granules [15]. A comparison of Raman signals and mechanical properties was made with silk fibres. The diameters of single degummed silk of domestic *Bombyx Mori* and wild silk of *Gonameta Rufobrunnea* (South-Africa) fibres are ~10 µm. Before examination in water, hair fibres were dipped a few hours in distilled water.

### Mechanical measurements

The mechanical properties of the samples were measured with an advanced Universal Fibre Tester, described elsewhere [14]. With an accuracy of ±0.01g on the monitored load and on the cross-head displacement to within ±0.001mm, we can assume very accurate mechanical measurements on single fibres. The specimen gauge length was 30mm. To reduce risks of failure in the grips, samples were stuck on a paper frame using neoprene glue. The different strain levels were applied at a strain rate of 100% min$^{-1}$. Mechanical tests were designed to take ~30 seconds. The effect of water absorption has been observed by placing a recipient of water under the fibre during loading (Figure 3). In this way the relative humidity (RH) could be raised to 100%.

*Raman spectrometer device*
An XY Raman microspectrometer (Dilor, Lille, France) equipped for signal detection with back-illuminated liquid nitrogen-cooled CCD matrices (2000×256 pixels) was used. Illumination was made with an Ar$^+$-Kr$^+$ ion laser. This instrument had a double holographic monochromator as a filter and the 514.531 nm excitation line was used (spectral resolution ~0.5 cm$^{-1}$) and its power density was kept to 3-4 mW µm$^{-2}$ measured before the sample, in order to avoid inducing any thermal effects in the keratin fibre structure. Backscattering illumination and collection of the scattered light were made trough the Olympus confocal microscope (long working distance Olympus ×50, total magnification ×500).

*Raman investigation under strain*
The combination of the Raman spectrometer and the Universal Fibre Tester allows the spectra to be recorded under macroscopic extension at different relative humidites (RH %). The laser was focused on the fibre core and was polarized either parallel to the fibre axis (noted //) or perpendicular (noted ⊥) by changing the alignment of the fibre with respect to the spectrometer slit. After stabilisation of the spatial dimension (relaxation) of the fibre (~300 s), the Raman spectra were recorded over 900 s (2 cycles; spectral range: 3100-3600 cm$^{-1}$), 1800 s (2 cycles, spectral range: 400-1800 cm$^{-1}$), and 900 s (low wavenumber: 0-400 cm$^{-1}$). To validate the spectral decomposition procedure, some recordings have been cycled 10-15 times, especially for the low wavenumber region. Such long recording times are required for natural fibres in order to extract Raman parameters with accuracy and a constant strain, instead of a constant stress considered in previous studies [11,13]. Contrary to dynamic tensile (macro)measurements, a complete Raman study of a keratin fibre extended at different strain to its failure would have taken several hours/days so that only few spectra were recorded for the five fibres studied.

*Data treatment*
Spectral baselines were subtracted using Labspec software (Dilor) and then decomposed using Origin software from Microcal Software (USA), according to the procedure previously described [11]. ttention was paid to the reproducibility of the data treatment respecting the Raman mode attributions found in the literature. Regarding this, the following assumption was made: Lorentzian shape for narrow peaks and a Gaussian form for broad peaks. The spectral evolutions and correlations were extracted using the laboratory-written CORDE software, which is based on a 2D correlation method [23,24]. Actually, nine perpendicular Raman spectra (1100-1800 cm$^{-1}$) with an increment of 5% strain were considered (0 → 40%).

## RESULTS

*Effect of water of macroscopic tensile properties*
Stress-strain curves for hair single fibres revealed two different behaviours depending on the degree of water exposure (Figure 4.a). At RH=40-60% (dry), a 3-regime plot was revealed, as observed for fibre-reinforced composites with weak fibre-matrix interfaces [25]: first a 0-4% strain linear elastic behaviour, then a 4-18% strain viscoelastic plateau, which is assigned to be an α helix → β sheet transition [9,10]. Finally, the curve exhibits a hardening regime up to the fracture. We notice that the transition takes place at constant stress, which supports well the assumption of a phase transition occurring. At RH= 100%, the curve of the hydrated fibre shows a continuous deformation requiring an increasing stress instead of the plateau. Moreover, wet hair exhibits a loss of initial stiffness ($E_i$ = 2.3 ± 0.1 GPa) and ultimate stress ($\sigma_R$ = 165 MPa) in comparison with dry hair ($E_i$ = 6.6 ± 0.1 GPa; $\sigma_R$ = 230 MPa). Relaxation measurements (Figure 4.b) were carried out up to ~30% strain. The recovery for hair deformed in water (ε = 8%) was greater than in a less humid environment (ε = 20%). With such dramatic differences in mechanical properties, the lubricating role of water in keratin fibres during extension is revealed.

*Choice of Raman probes*
An overview of Raman spectra of synthetic and natural polymeric fibres such as PA-66 [11], silk [26] and hair (this work) should be considered (Fig. 5). They all contain some amide groups (CONH) and thus exhibit common Raman modes, such as amide I & III, or –N-H stretching mode (high frequencies) visible mainly in the perpendicular polarization. Note the more complex Raman signal for natural fibres in comparison with PA-66, which reveals the characteristics of a hierarchical and less organized material. Complex proteins sequences induce necessarily specific local conformation and hence orientation disorder. In analogy with previous studies on the amide-based fibres [11,12,26], we must use the modes which are considered here as mechanical probes for deformation of the nano-structure (e.g. the interchain arrangements). Indeed, the amide I mode is linked to the secondary structure of proteins, and the –N-H stretching mode can be used a direct measure of hydrogen bonds distance (intra/inter) of keratin chains. Contrary to silk and PA-66, keratin macromolecular chain shows the –S-S- stretching mode around 500cm$^{-1}$ on its Raman spectra which could be revealing the contribution of disulphide bonds to the macromolecule coupling. As keratin fibres exhibit a complex signal, it is necessary that the spectral decomposition method takes rigorously into account the evolutions of Raman modes.

*Validation of spectral decomposition*
Spectra recording of keratin fibres were carried out for various levels of extension. As mentioned by Kuzuhara [15-18], intensity normalizations were made using the phenylalanine peak (very narrow 1003cm$^{-1}$ peak, characteristic of an isolated vibrating entity) for the spectral range 400-1100cm$^{-1}$ or the CH$_2$ scissoring band (1450cm$^{-1}$) on the range 1100-1800cm$^{-1}$. Note that we initially used a neon lamp (peak at 916cm$^{-1}$ using green excitation and corresponding Raman shift) before checking that the phenylalanine peak consisted of an efficient reference as well. To establish a convenient fitting procedure, 10-12 cycles recording spectra were first considered. Once spectral decompositions had been reproduced successfully with the increasing strain (0→45%), a consistent evolution of the "fitted" bands was observed depending both on the polarization and the relative humidity (Figs 6 & 8). For instance, the amide I mode (1650-1680cm$^{-1}$), decomposed with a Lorentzian (α helix) and a Gaussian (β sheet), revealed the assumed α → β transition at least from 4 % strain (Fig. 7). Indeed, a continuous increase of the β sheet component (conversely for the α helix component) was observed. Thus, the chosen descriptions could be used to extract data concerning wavenumber shifts or intensity evolutions but an independent method without any arbitrary role regarding the setting of the components (number, position and type of bands) is required to confirm the validity of the fit. This is supplied by the use of the 2D correlation technique (C2D), introduced notably in vibrational spectroscopy by Noda [23]. As seen previously on SiC fibres [24], a halo on the C2D map corresponds to a correlated evolution in intensity. Hence, the synchronous C2D treatment on the range 1100-1800cm$^{-1}$ (Fig. 8) gives an obvious agreement for the existence of Raman modes changing linearly in opposite ways (opposite colours / seen on web issue) towards 1600-1700cm$^{-1}$. Focusing on the range 1500-1800cm$^{-1}$, this correlation is characterized by two halos on the diagonal (intensity variations) attributed by their well-defined position to the amide I mode in the α helix configuration (1650cm$^{-1}$) and the β sheet configuration (1670-1690cm$^{-1}$). Consequently both treatments lead to similar results regarding the existence of two independent components in the amide I mode. Finally, the C2D technique confirms the bands set into the spectral decompositions as we have demonstrated in the range 1500-1800cm$^{-1}$.

## *Raman analysis under extension*

### *Inter-chain probe*

Focusing on the -N-H stretching mode (3100-3700cm$^{-1}$), information about the modification of neighbouring chains arrangements was analysed. Introducing four bands in the "fit" (Fig. 9), we aim to represent the well-organized (crystalline) part (Lorentzian: L) as well as the amorphous regions of the material (Gaussian) which contributes to the mode. Note the increasing importance of the Gaussian band at 3415cm$^{-1}$ with RH. This component corresponds to the stretching vibration mode -O-H increased by the presence of water [26]. As shown in figure 10, the behaviour of the bands must be differentiated depending of the level of humidity. At RH ~ 40-60%, the crystalline (L) component, shifts towards higher wavenumbers, for 0-4 % strain, and then exhibits a plateau until the fracture (Figure 10.a). At RH = 100%, despite only few data, the opposite behaviour (a decrease) is observed. The amorphous component (G) also shifts ,towards higher (lower) wavenumbers at RH ~ 40-60% (RH =100%) but without showing any regime (Fig. 10.b). As a correlation exists between –N-H stretching wavenumber and the –X-H⋯Y distance [27,28], we must interpret the wavenumber upshift of the lorentzian component as an increase of the hydrogen bonding distance in the crystalline part (IFs) until 4% strain. At 4%, the α → β transition starts, α helices unfold (Figure 12) and make hydrogen bondings between segments, chains, nor dimmer (Fig. 11). That should explain the plateau. Conversely, the presence of water involves a decrease or at least a constant –N-H⋯O distance. Both cases can be explained: water saturates most of the hydrogen bonds instead of the keratin molecules, so an applied strain does not take into account the increasing distance between keratin chains and no effect on the –N-H mode. The shift towards low wavenumbers may be interpreted by the squeezing out of water between adjacent chains (Poisson' effect), which would reinforce the hydrogen bonds and so decreases the wavenumber.

The –S-S- stretching mode is representative of the disulphides bonds which link two adjacent keratin molecules. After normalization of the Raman spectra with the phenylalanine peak, the –S-S- peak is used to assess the amount of disulphide bonds: the intensity ratio $I_{S-S}/I_{Phe}$ exhibits a significant decrease (-20%) between 0 and ~4% strain, then a plateau, and up to ~20 % decreases again until fracture in dry air (Fig. 13.a). As similar behaviour was observed in parallel polarization (not shown here), it is obvious that a large number of disulphide bonds are broken during tensile loading.

### *Intra-chain probe*

The stretching CONH mode (amide I) is closely linked to the secondary structure (e.g. α helix). At RH ~ 40-60%, an increase in wavenumber is observed from 0 to 4% strain, indicating a compression (+2cm$^{-1}$) of the CONH segment (α helix structure) which is attributed to the Poisson' effect. During the α → β transition, the amide I mode is still compressed around 1652cm$^{-1}$ and corresponds to a plateau between 4 and ~20 %. Up to 20 %, a down-shift appears until fracture: this indicates that the macromolecular (crystalline) chains undergo an expansion under the action of the macroscopic tension up to the fracture. Saturated with water, a different behaviour is observed: as shown on the Fig. 13.b, there is no compression but a plateau until ~20 %, leading to fractureand down-shift.

## DISCUSSION

*Nano & macromechanics correlation*
As shown on Fig. 13, the Raman probes describe the different macromechanical steps. The Raman measurements of the different probes match obviously with the macromechanical behaviour, both in dry air and water and thus comprise a unique tool describing, at the nanometre scale, the macromolecular deformation occurring during the macroscopic strain of the fibre.

*Role of water*
There is a combination of two phenomena resulting from the macroscopic strain: molecular stretching and sliding of adjacent macromolecules facilitated by the presence of water which saturates the hydrogen bonding sites (Figure 13.b). Indeed, water intercalation spreads out the α helix → β sheet transition as seen on Fig. 7 and modifies the interaction between probed "crystalline" keratine fibres and their environment. This is consistent with a decrease of the phase-transition threshold and hence of the "fibre-matrix" interface bond (composite description) and of the inter-chain links.

*Modelling the nanostructure*
Mechanical models have been established for keratin fibres both by Feughelman (F94) and by Chapman & Hearle (C/H). They consider either some links between IFs, consisting of disulphide bonds situated mostly in the terminal domains, but they diverge radically on the assumption of linkage between the two parts of the macrofibrils, as well as the IFs and the amorphous matrix [1,2]. For Feughelman [1], the two parts are totally independent, and conversely for Chapman & Hearle [2]. It evident, that the required energy to change bond angles of a chemical system is far less than that involving bond breakage. Thanks to Figs 7 and 8, it can be seen that a α helix → β transition occurs at ~4% strain and probably continues up to fracture for amorphous macromolecular chains. The (instant) phase transition takes place only in the crystallized part (IFs) and is easier to make breaking the disulphide bonds between terminal domains of the IFs. However, Figure 13.a gives a proof of the rupture of disulphides bonds from the start of straining (0-4 %). This suggests the existence and thus the rupture of the linkages between the matrix and the IFs. Consequently,the model of Chapman & Hearle is seen to be more realistic.

In previous studies of advanced synthetic fibres [11,13] it has been possibleto discriminate between the low wavenumber signatures of crystalline and amorphous macromolecular chains because of their high crystallinity and high anisotropy. Fig. 14 compares polarized spectra recorded for dry hair fibres and the evolution of the main parameter as a function of the fibre strain. The weaker contribution of the Lorentzian components indicates a low crystallinity. A rather strong axial character can be recognized from the parallel/perpendicular polarized spectra. A significant loss of polarization is obvious as a function of the strain. As a guide a straight line is drawn, but most of the decrease seems take place before the phase transition

(4% threshold). Comparison with low temperature spectra is required to asses our description more precisely and to deeper in the understanding of the mechanics of keratine fibres.

## ACKNOWLEDGMENTS

D. Baron is acknowledged for developing the CORDE software. The authors thanks Dr A. Bunsell for many discussion and the critical reading of the manuscript.

FIGURE CAPTIONS

**Fig. 1.** Hierarchical structure of hair. Intra- and inter-hydrogen (dash) and covalent disulphide (-S-S-) bonds (yellow balls)) are expected to link keratin molecules (α helix).

**Fig. 2.** Stress-strain relationships for hair (Caucasian) and wild silk (Gonameta rufobrunnea) single fibres at ~40-60% relative humidity (RH). Comparison is made with PET, PA-66 and i-PP fibres at RH = 50 (±5) % and correlation between ultimate strength and density is given.

**Fig. 3.** Detail of the experimental setting showing the focused laser beam on the single filament fibre (arrow) under controlled stress/strain. A 100% relative humidity environment can be reproduced using the meniscus forms by an exceeding drop of water in which the fibre goes through.

**Fig. 4.** (a) Stress-strain curves for hair single fibres at RH ~ 40-60% and 100%. A rectangle shows the 4-18% strain at constant stress corresponding to α helix →β sheet transition. (b) Mechanical hysteresis loops between 0 and 30% strain at RH =40-60% and 100%.

**Fig. 5.** Comparison of parallel (400-1800$cm^{-1}$) and perpendicular (3100-3800$cm^{-1}$) polarised Raman spectra of hair, silk and PA-66 fibres.

**Fig. 6.** Example of spectra recorded for different level of tensile strain at RH ~ 40-60 % (a; b) and RH = 100% (c). Gaussian and Lorentzian components are shown in the procedure used to extract wavenumber shifts and intensity evolutions occurring for Raman vibration modes in the 400-1800$cm^{-1}$ range for parallel (//) and perpendicular ($\perp$) polarisation.

**Fig. 7.** The perpendicular polarised CONH stretching mode has been decomposed by a Lorentzian and a Gaussian component respectively associated to α helix and β sheet component. The α helix configuration plotted as a function of strain decreases giving an increase of the β sheet configuration at RH ~ 40-60% and 100%.

**Fig. 8.** Synchronous 2D-correlation components from a series of nine perpendicular polarised Raman spectra of hair under 5% incremental strains in the 1180-1800$cm^{-1}$ range. Details of the 1500-1800$cm^{-1}$ region are given.

**Fig. 9.** Procedure used to extract wavenumber shifts and intensity evolutions occurring for the –N-H stretching mode ($\nu_{-N-H}$) in the 3100-3800$cm^{-1}$ range for perpendicular ($\perp$) polarisation under loading at RH ~ 40-60% (a) and 100% (b). Amorphous (G) and Crystalline (L) components have been deduced. The contribution of water (G($H_2$0)) has been taken into account.

**Fig. 10.** Raman wavenumber shifts of the $\nu_{-N-H}$ mode plotted as a function of increasing applied strain at RH ~ 40-60% and 100%. Crystalline (a) and amorphous (b) components are given.

**Fig. 11.** Schematic representation of the keratin intra and inter-chain/dimmer hydrogen bonds. Hydrogen bonding/dimmer distance and helical/dimmer repeat unit lengths are given according to X-Ray Diffraction studies [4-8].

**Fig. 12.** Molecular scheme of the α helix keratin before (left) and after extension (right).

**Fig. 13.** Wavenumber shifts of the $\nu_{-N-H}$ (crystalline) and the $\nu_{-CONH}$ (α helix) amide I mode and decreasing of the –S-S- Raman signature (the ratio of the peak intensity = -S-S- band/Phe band) plotted as a function of increasing applied strain at RH ~ 40-60% (a) and in water (b). In both cases, a comparison is made with macroscopic strain/stress behaviour.

**Fig. 14.** Low wavenumber parallel and perpendicular Raman spectra (RH ~ 40-60%). Amorphous (Gaussian: G) and crystalline (Lorenztian: L) components have been deduced. A significant loss of polarization parallel to the fibre axis part is shown for the crystalline part (b) as contrary for the amorphous one (c).

Table 1 : Main wavenumbers (in bold, fitted) assignments after 15-24 and references herein.

| wavenumber / cm$^{-1}$ | Assignments |
|---|---|
| 422 | δ(CCC) chain |
| **506 (L)** | ν(SS) [trans-gauche-trans] |
| 529 | ν(SS) [gauche-gauche-trans] |
| 540 | ν(SS) [gauche-gauche-gauche] |
| **600 (L)** | ρ(CH) wagging |
| **627 (L)** | ν(CS) [gauche] |
| **643 (L)** | ν(CS) [gauche] |
| **741 (G)** | ρ(CH$_2$) |
| **751 (L)** | ρ(CH$_2$) |
| **830 (L)** | δ(CCH) aliphatic/ *tyrosine* |
| **855 (L)** | δ(CCH) aromatic/ *tyrosine* |
| **880 (L)** | ρ(CH$_2$)/ν(CC)/ν(CN)/*Tryptophane* |
| **900 (L)** | ρ(CH$_2$) |
| **934 (L)** | ν(CC) helix α/ ρ(CH$_3$) terminal |
| **1004 (L)** | ν(CC) ring/*Phénylalanine* |
| **1033 (L)** | ν(CC) skelettal/*cis* |
| 1061 | ν(CC) skeletal/*trans* |
| **1081 (G)** | ν(CC) skeletal/*random* |
| 1102 | ν(CC) skeletal/ *trans* |
| **1128 (G)** | ν(CC) skeletal/ *trans* |
| **1161 (L)** | ν(CC)/δ(COH) |
| **1177 (L)** | ν(CC) |
| 1185 | ν(CC) |
| **1209 (L)** | ν(CC)/*Tyrosine/Phenylalanine/amide III** |
| **1239 (L)** | δ(CH$_2$) wagging/ν(CN)/ *amide III disordered* |
| 1273 | ν(CN)/δ(NH)/ *amide III (α helix)* |
| **1301 (L)** | δ(CH$_2$) |
| **1315 (L)** | δ(CH$_2$) |
| 1391 | δ(CH$_3$) symetric |
| 1400 | δ[(CH$_3$)$_2$] symmetric |
| 1425 | δ(CH$_3$) |
| **1452 (L)** | δ(CH$_2$) scissoring (*lipid/protein*) |
| 1534 | δ(NH) |
| **1556 (L)** | δ(NH)/ν(CN)/*amide II* |
| **1586 (L)** | ν(C=C) ring |
| 1609 | *Phenylalanine/Tyrosine* |
| **1616 (L)** | ν(C=C) ring/ *Tyrosine/Tryptophane* |
| **1654 (L)** | ν(CONH) amide I α helix |
| **1677 (G)** | ν(CONH) amide I β sheet |
| *2565 (ns)* | ν(SH) |
| *2732 (ns)* | ν(CH) aliphatic |
| *2875 (ns)* | ν(CH$_2$) symmetric |
| *2931 (ns)* | ν(CH$_3$) symmetric |
| *2966 (ns)* | ν(CH$_3$) asymetric |
| *3060 (ns)* | ν(CH) ring |
| **3284 (L)** | ν(NH), *crystalline* |
| **3325 (G)** | ν(NH), *crystalline* |
| **3415 (G)** | νH$_2$O |

ns : not studied; L : Lorantzian; G: Gaussian

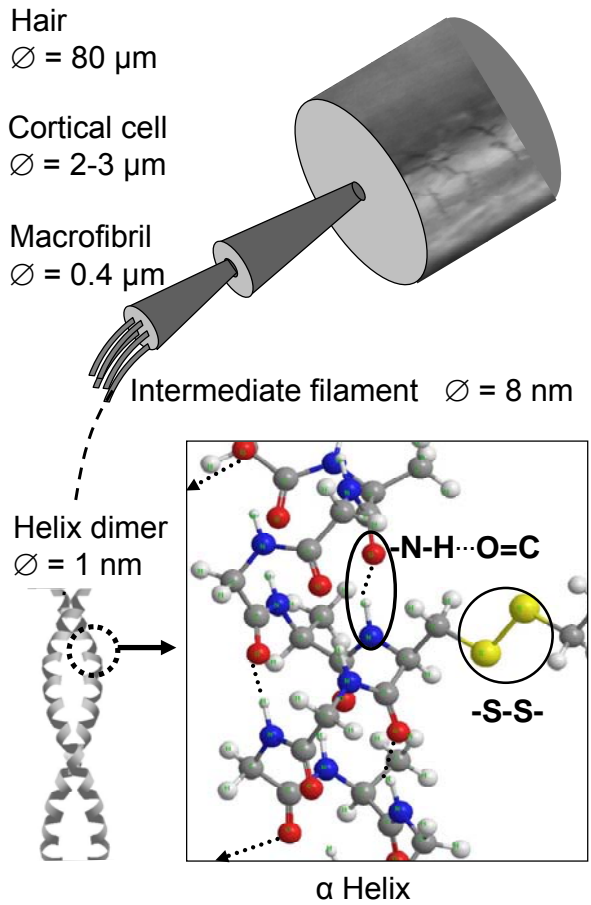

**Fig. 1.**

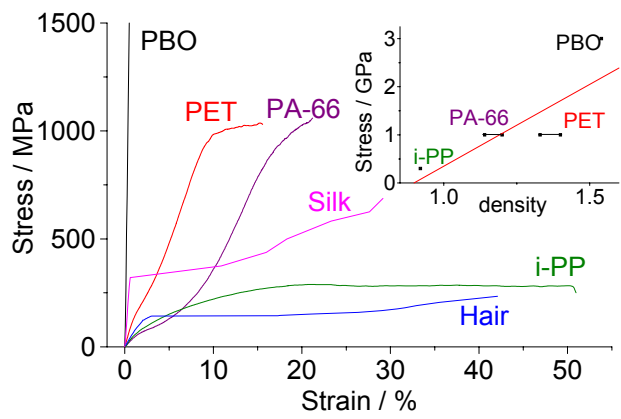

**Fig. 2.**

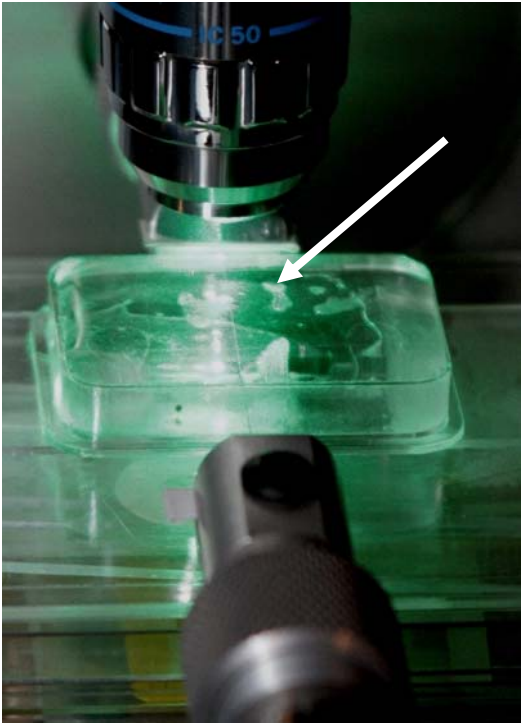

**Fig. 3.**

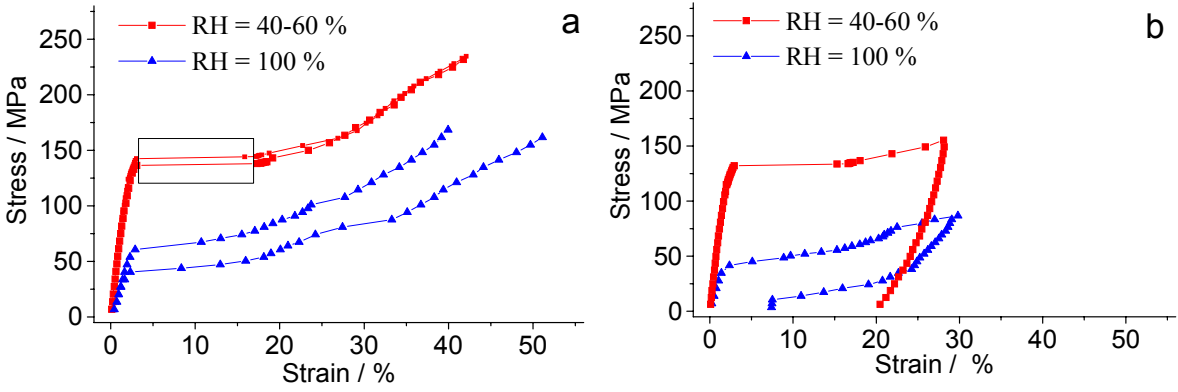

**Fig. 4.**

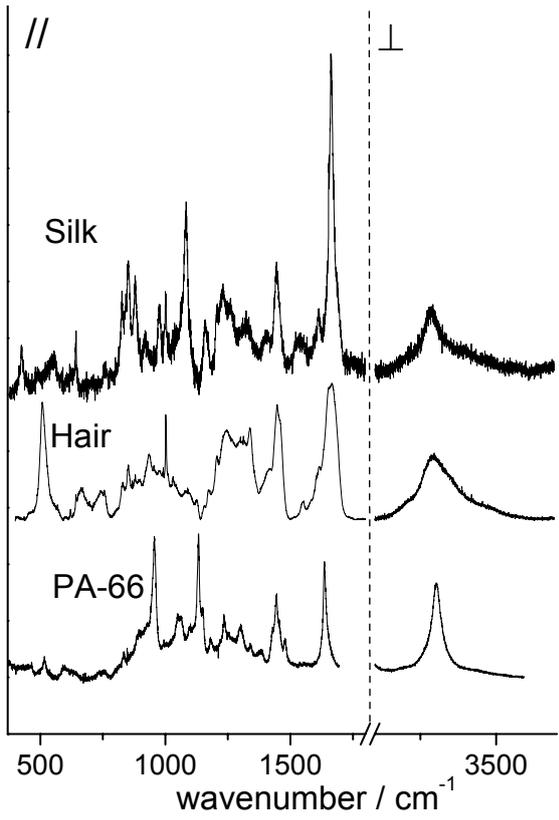

**Fig. 5.**

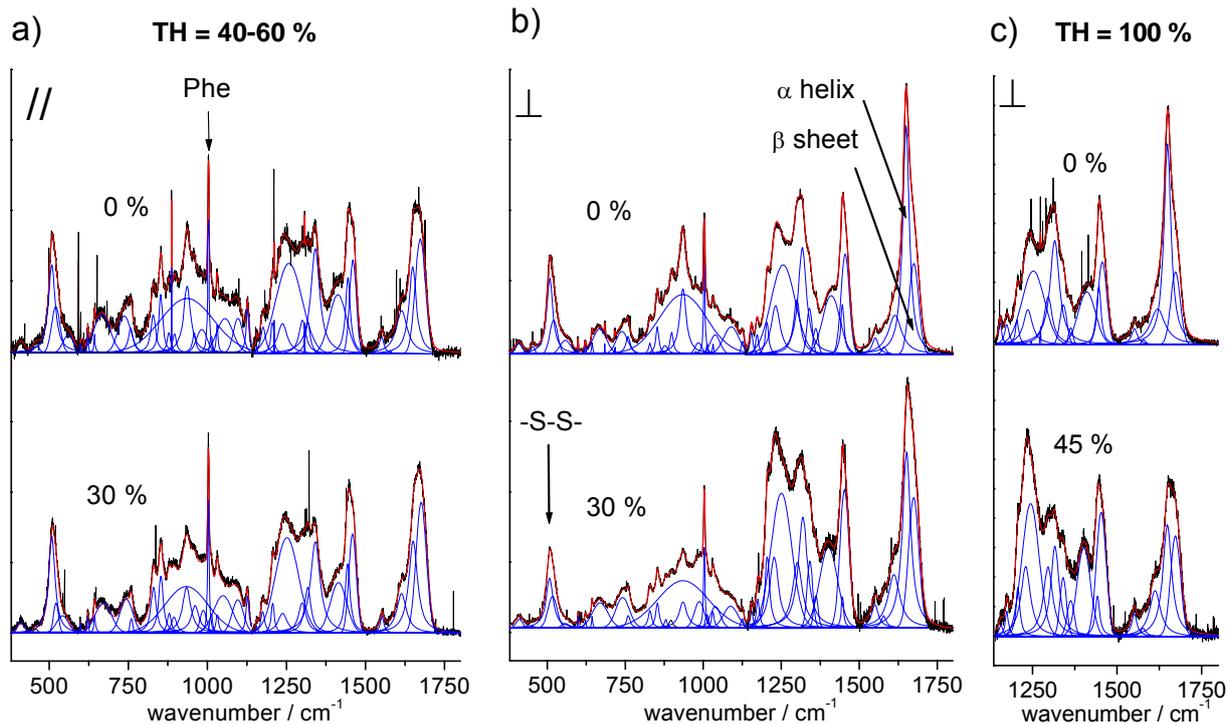

**Fig. 6.**

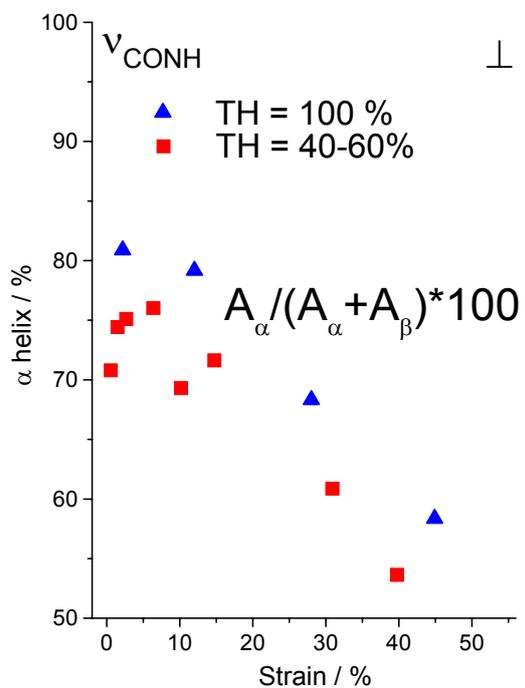

**Fig. 7.**

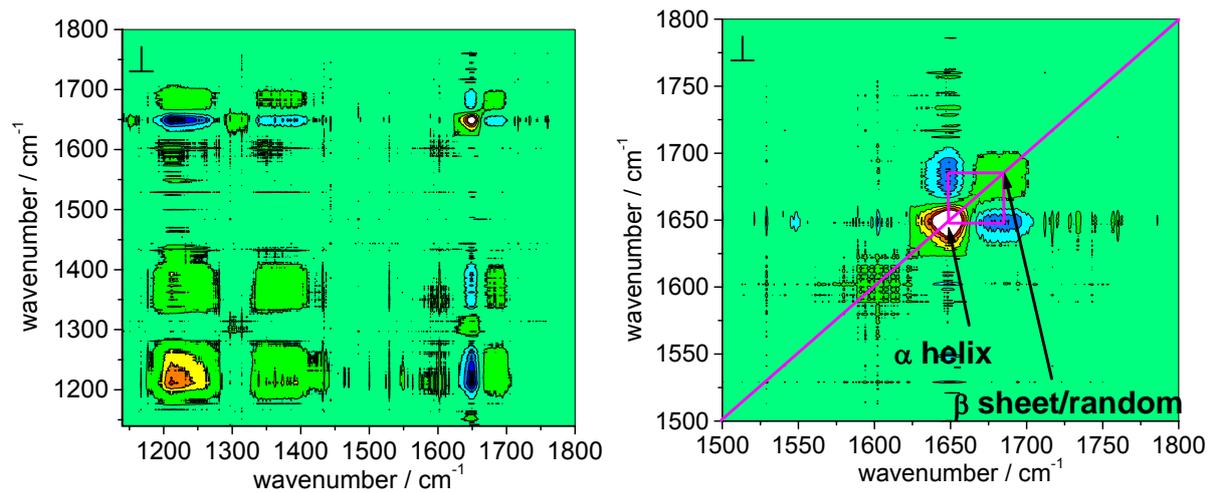

**Fig. 8.**

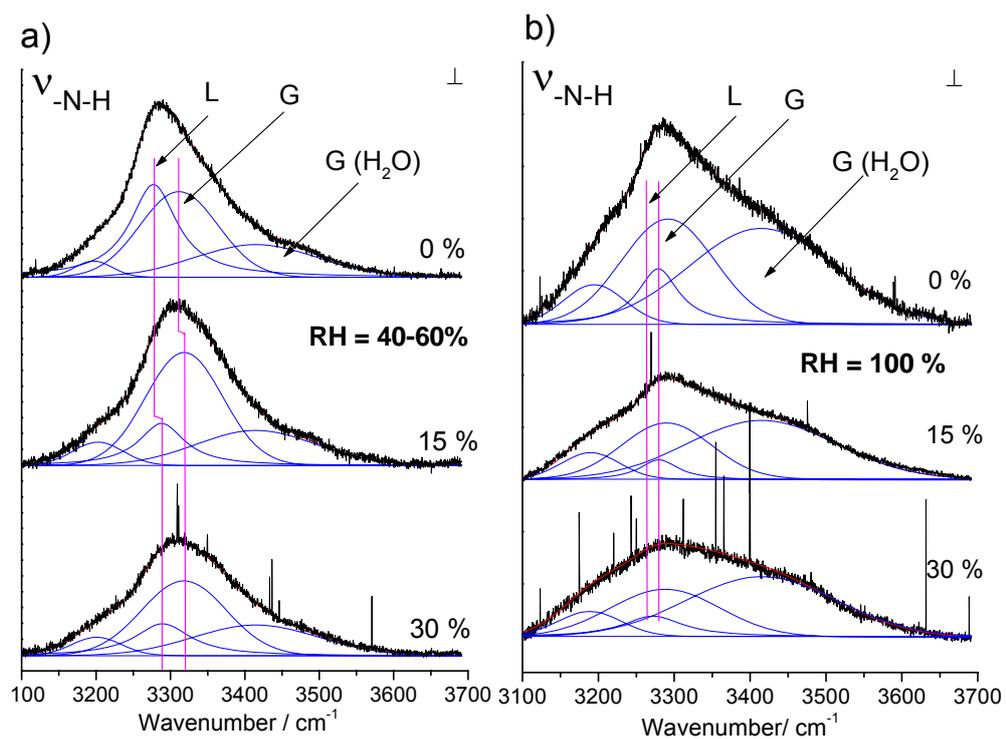

**Fig. 9.**

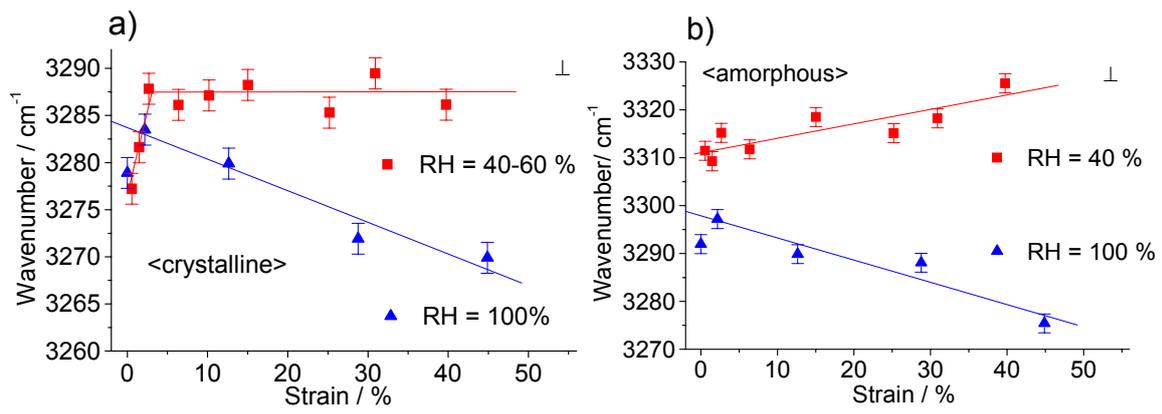

**Fig. 10.**

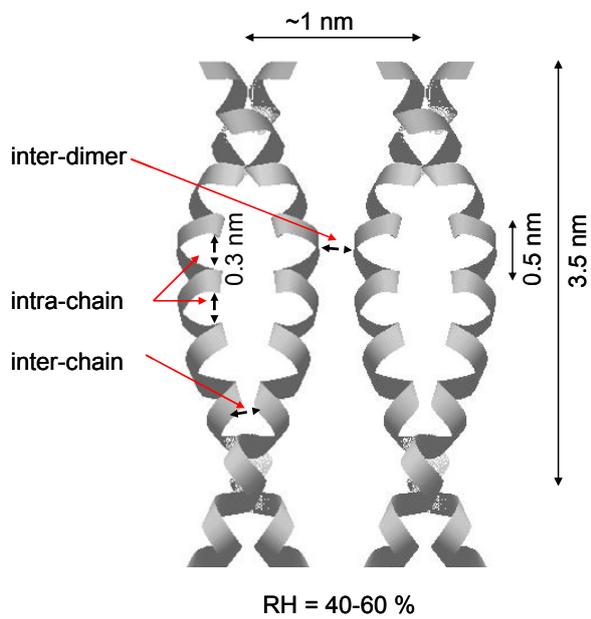

RH = 40-60 %

**Fig. 11.**

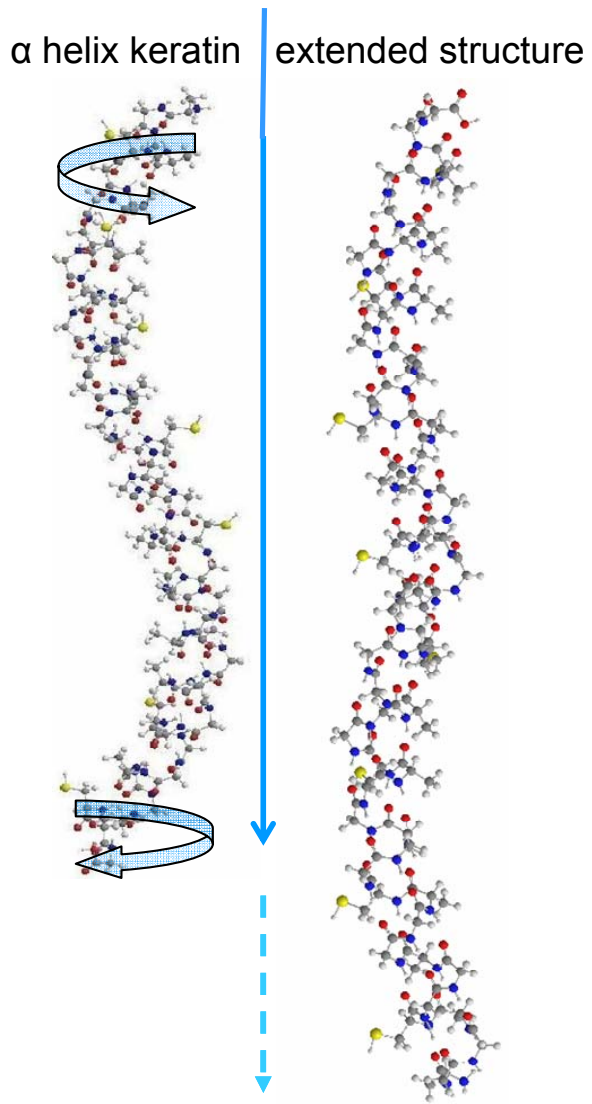

**Fig. 12.**

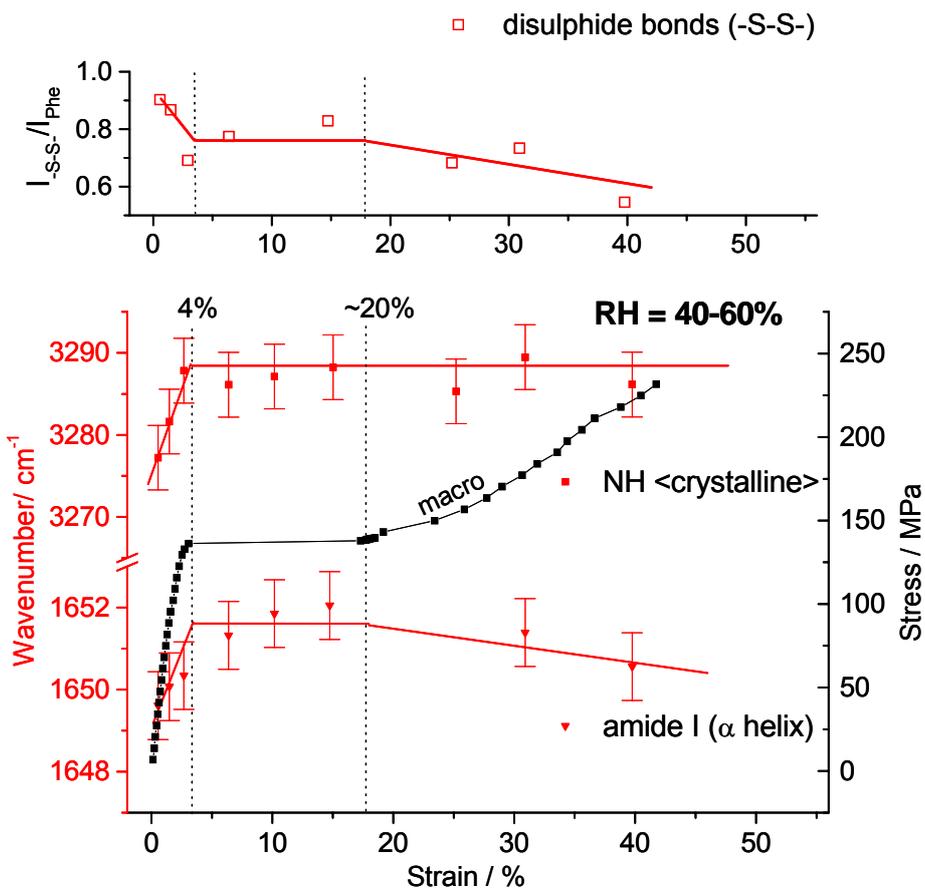

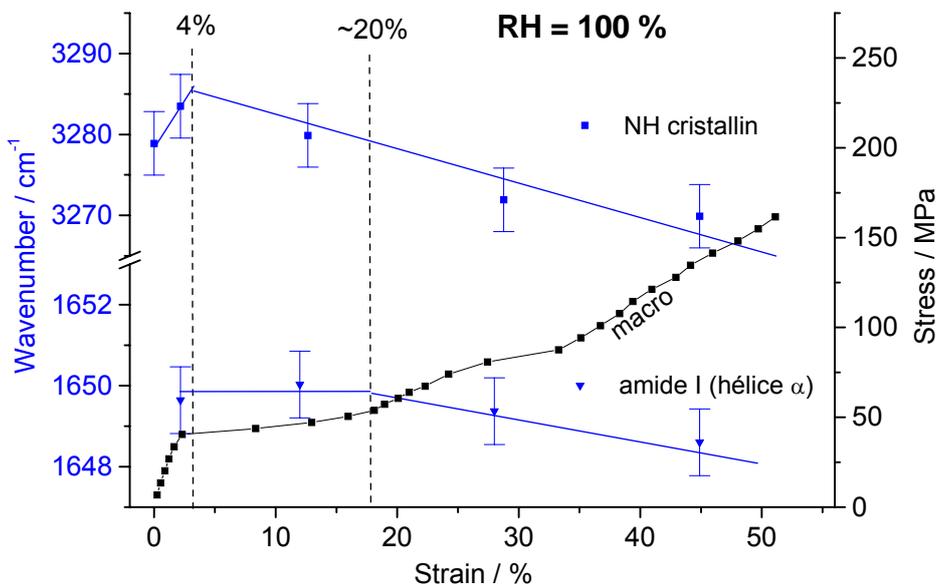

**Fig. 13.**

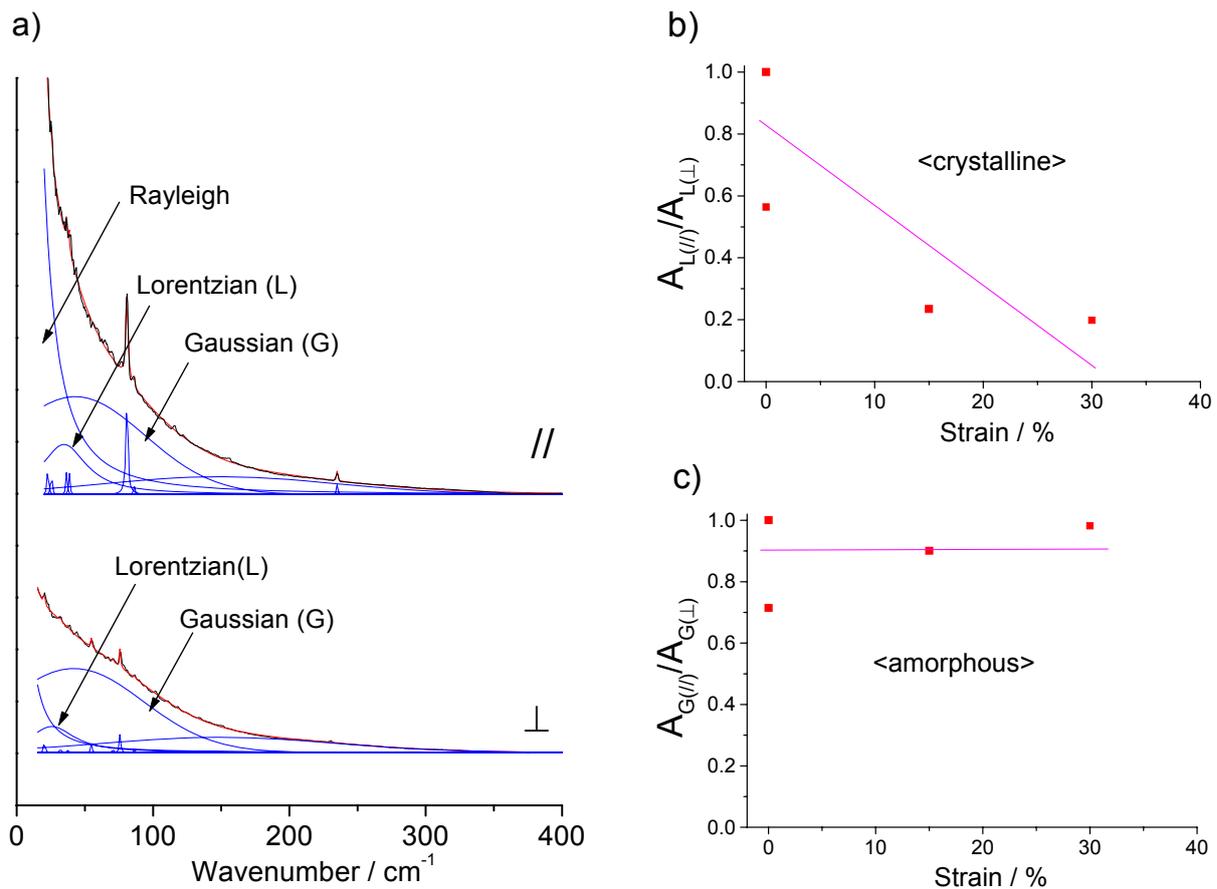

**Fig. 14.**